\documentclass[a4paper,10pt,twoside]{cpc-hepnp}

\usepackage{multicol}
\usepackage{graphicx}
\usepackage{booktabs}
\usepackage{amssymb,bm,mathrsfs,bbm,amscd}
\usepackage[tbtags]{amsmath}
\usepackage{lastpage}
\usepackage{CJK}
\usepackage{subfigure}
\usepackage{color}

\newcommand{\jpsi}{J/\psi}

\newcommand{\pip}{\pi^+}
\newcommand{\pim}{\pi^-}

\newcommand{\pbar}{\overline p}

\newcommand{\pppipi}{p \overline p \pi^+ \pi^-}

\begin{document}

\fancyhead[c]{\small Chinese Physics C~~~Vol. XX, No. X (201X)
XXXXXX} \fancyfoot[C]{\small 010201-\thepage}

\footnotetext[0]{Accepted by Chinese Physics C}

\title{Study of tracking efficiency and its systematic uncertainty from $\jpsi \to \pppipi$ at BESIII\thanks{Supported by Joint Funds of National Natural Science Foundation of China(U1232201), National Natural Science Foundation of China (11275210, 11205182, 11205184), and National Key Basic Research Program of China (2015CB856700). }}

\author{%
      Wen-Long Yuan$^{1;1)}$\email{yuanwl@ihep.ac.cn}%
\quad Xiao-Cong Ai$^{2}$
\quad Xiao-Bin Ji$^{2}$\\
\quad Shen-Jian Chen$^{1}$
\quad Yao Zhang$^{2}$
\quad Ling-Hui Wu$^{2}$\\
\quad Liang-Liang Wang$^{2}$
\quad Ye Yuan$^{2}$
}
\maketitle

\address{%
$^1$ Nanjing University, Nanjing 210093, China\\
$^2$ Institute of High Energy Physics, Chinese Academy of Sciences, Beijing 100049, China\\
}

\begin{abstract}
Based on $J/\psi$ events collected with the BESIII detector, with corresponding Monte Carlo samples, the tracking efficiency and its systematic uncertainty are studied using a control sample of $\jpsi \to \pppipi$. Validation methods and different factors influencing  the tracking efficiency are presented in detail. The tracking efficiency and its systematic uncertainty for protons and pions with the transverse momentum and polar angle dependence are also discussed.
\end{abstract}



\footnotetext[0]{\hspace*{-3mm}\raisebox{0.3ex}{$\scriptstyle\copyright$}2013
Chinese Physical Society and the Institute of High Energy Physics
of the Chinese Academy of Sciences and the Institute
of Modern Physics of the Chinese Academy of Sciences and IOP Publishing Ltd}%

\begin{multicols}{2}

\section{Introduction}

The charged track reconstruction (tracking) efficiency and its systematic uncertainty play an important role in high energy physics experiments. A detailed study to develop a robust methodology for determining the tracking efficiency is essential for both software development and experimental data analysis. However, the tracking efficiency is not always given a good definition or straightforward determination. This paper, focusing on the methodology, presents a detailed study of tracking efficiency and its systematic uncertainty, with various influencing factors on the efficiency, as well as a number of essential validations.

The key tracking sub-detector in the BESIII detector ~\cite{bes3} at the Beijing Electron-Positron Collider (BEPCII)~\cite{bepc} is a small-celled, helium-based main drift chamber (MDC) with 43 layers of wires, which has a geometrical acceptance of $93\%$ of $4\pi$ and provides momentum and $dE/dx$ measurements of charged particles. The other components of BESIII mainly include an electromagnetic calorimeter (EMC) made of CsI(Tl) crystals, a plastic scintillator time-of-flight system (TOF), a super-conducting solenoid magnet, and a muon chamber system (MUC) made of Resistive Plate Chambers (RPCs).

A clean sample of $\jpsi \to \pppipi$ decays is chosen for this study, due to low background level, high statistics, broad momentum and angular distributions, etc. A data sample of $(255.3\pm2.8) \times 10^6 \jpsi$ events collected with BESIII is used here.~\cite{jpsi} The optimization of the event selection, the investigation of possible backgrounds, and many validations in this study are performed using Monte Carlo (MC) simulated data samples. The {\sc geant4}-based simulation software {\sc{boost}}~\cite{boost} includes the geometric and material description of the BESIII detector and the detector response and digitization models, as well as the tracking of the detector running conditions and performance. The production of the $\jpsi$ resonance is simulated by the MC event generator {\sc kkmc}~\cite{kkmc}; the known decay modes are generated by {\sc evtgen}~\cite{evtgen} with branching ratios set at Particle Data Group (PDG) values\cite{pdg}, while the remaining unknown decay modes are modeled by {\sc lundcharm}~\cite{lundcharm}.

\section{Definition}
The tracking efficiency ($\epsilon_{trk}$) for a given particle is defined as
\begin{equation}
\epsilon_{trk}=\frac{n}{N}
\label{def_tracking}
\end{equation}
where the denominator ($N$) is the number of signal events with all other particles in the final states required to be reconstructed except the one under study, no matter whether the studied particle is reconstructed or not; the numerator ($n$) is the number of events with all particles in the final states reconstructed, including the one under study.

For example, in the calculation of the tracking efficiency for protons with the control sample  $\jpsi \to \pppipi$, at least three charged tracks should be reconstructed and identified as $\pbar$, $\pip$, and $\pim$, respectively. The number of signal events ($N$) is obtained by fitting the missing mass distribution of $\pbar$, $\pip$, and $\pim$. The missing mass is defined as $M_{miss} = \sqrt{E_{miss}^{2} - \vec P_{miss}^{2}}$, where the missing four-momentum $(E_{miss}, \vec P_{miss})$ is determined from the difference between the net four-momentum of the $\jpsi$ particle and the sum of the four-momenta of $\pbar, \pip,$ and $\pim$ in the event. The numerator ($n$) is  obtained by fitting the missing mass distribution with the fourth charged track reconstructed.

It is an advantage to obtain both $n$ and $N$ by fitting the missing mass distribution, since one can use the same function to describe the signal and  part of the errors in the fitting can be cancelled.

The systematic uncertainty of tracking ($\Delta_{trk}$) is defined as the difference of  tracking efficiency between MC ($\epsilon_{trk}(MC)$) and data ($\epsilon_{trk}(data)$)

Two hundred million inclusive $J/\psi$ MC events were used to investigate the possible backgrounds from $J/\psi$ decays.  Table \ref{pppipi_eff_bkg_ana} shows a topology analysis of the inclusive MC sample after event selection which assigns the proton as the missing particle. The intermediate resonances to the same $\pppipi$ final state have been treated as signals in the tracking efficiency calculation, except those with long-lived intermediate states, such as the $\Lambda$ particle. The tight vertex cut used here can result in a lower efficiency for long-lived particles, e.g. the tracking efficiency may lower by about $15\%$ at low momentum when including the $\Lambda$ particle and applying the same selection criteria.  The total background ratio is less than $0.5\%$ in a $3 \sigma$ signal region ($\sigma$ is the missing mass resolution), and a similar result can be obtained when assigning the missing particle to $\overline{p},\pi^+$, or $\pi^-$, respectively.

\begin{equation}
\Delta_{trk}=1-\frac{{\epsilon_{trk}(MC)}}{{\epsilon_{trk}(data)}}
\label{def_tracking_sys}
\end{equation}

Since $n$ is a subset of $N$, the correlation between $n$ and $N$ is $n$, and the covariance matrix of $(n, N)$ will be in the form:
\begin{equation}
\begin{split}
cov(n, n) = (\sigma_{n})^2,~~cov(N, N) = (\sigma_{N})^2,\\
cov(n, N) = (\sigma_{n})^2,~~cov(N, n) = (\sigma_{n})^2~
\label{def_cov}
\end{split}
\end{equation}
with the derivative
\begin{equation}
\frac{\partial\epsilon_{trk}}{\partial N} = -\frac{\epsilon_{trk}}{N},~~\frac{\partial\epsilon_{trk}}{\partial n} = \frac{1}{N}
\label{def_derivative}
\end{equation}
where $\sigma_{n}$ and $\sigma_{N}$ are given by the statistical errors of $n$ and $N$, respectively.

The error on the tracking efficiency ($\sigma_{\epsilon_{trk}}$) is
\begin{equation}
\begin{split}
\sigma_{\epsilon_{trk}} = \sqrt{(-\frac{\epsilon_{trk}}{N}~\frac{1}{N}) \left(\begin{array}{cc}(\sigma_{N})^2&(\sigma_{n})^2\\(\sigma_{n})^2&(\sigma_{n})^2\end{array}\right) \binom{-\frac{\epsilon_{trk}}{N}}{\frac{1}{N}}}\\ = \frac{1}{N}\sqrt{(1-2\epsilon_{trk})(\sigma_{n})^2 + \epsilon_{trk}^2(\sigma_{N})^2}~~~~~~~~~~~~~
\end{split}
\label{def_trk_err}
\end{equation}

In Eq.~(\ref{def_tracking_sys}), due to the independence of $\epsilon_{trk}(MC)$ and $\epsilon_{trk}(data)$, the error of the systematic uncertainty of tracking ($\sigma_{\Delta_{trk}}$) is
\begin{equation}
\sigma_{\Delta_{trk}}=(1 - \Delta_{trk}) \cdot \sqrt{\frac{\sigma^2_{\epsilon_{trk}(MC)}
}{\epsilon^2_{trk}(MC)}+\frac{\sigma^2_{\epsilon_{trk}(data)}
}{\epsilon^2_{trk}(data)}}
 \label{def_tracking_sys_err}
\end{equation}

\end{multicols}
\begin{center}
\begin{figure}
  \centering
  \subfigure{
    \includegraphics[scale=0.35]{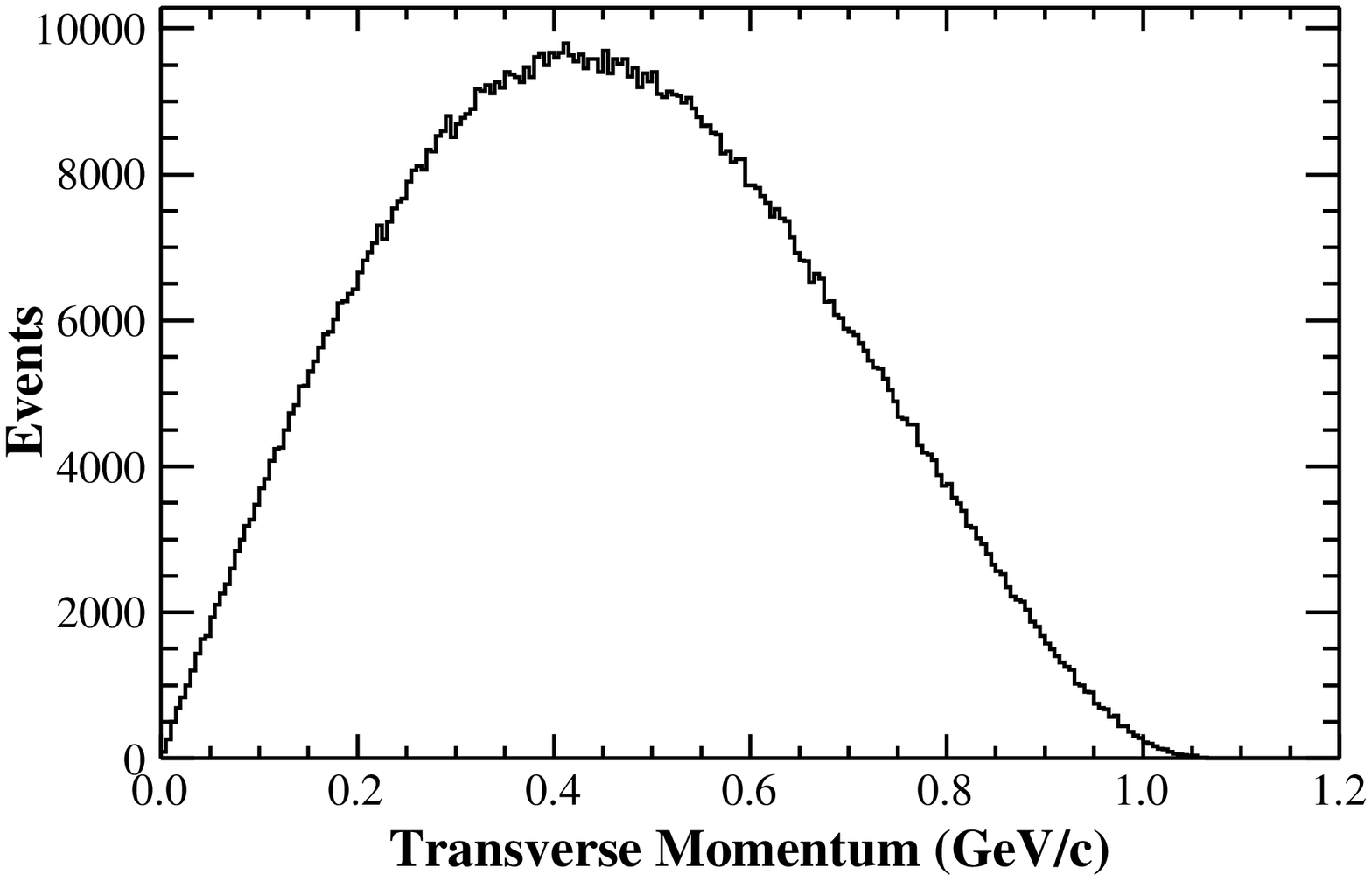}
    \put(-40,110){\bf (a)}}
  \subfigure{
    \includegraphics[scale=0.35]{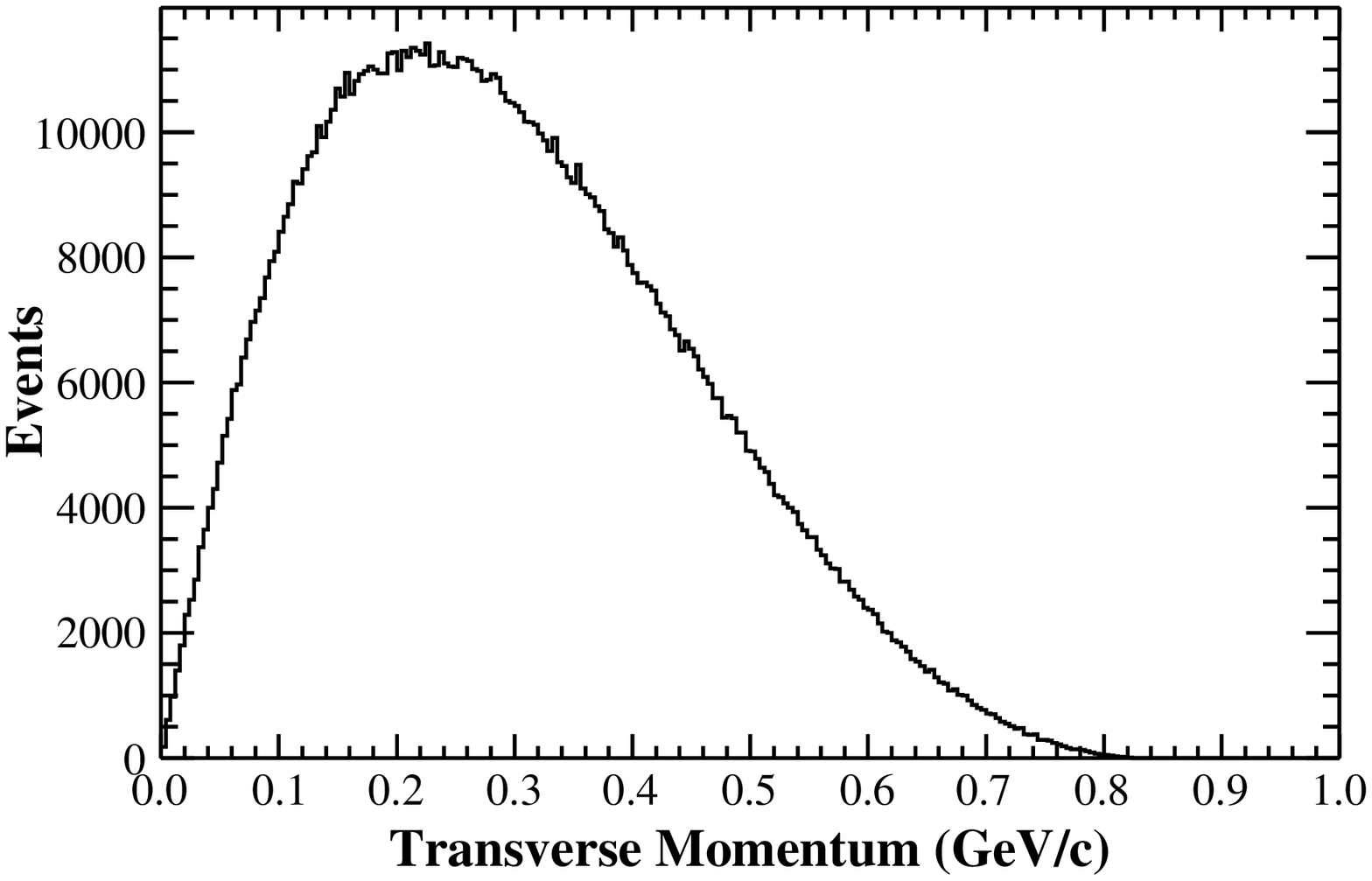}
    \put(-40,110){\bf (b)}}
    \figcaption{\label{pt_dis} Typical $P_T$ distribution of (a) protons and (b) pions from exclusive MC sample.}
\end{figure}
\end{center}

\begin{multicols}{2}

\section{\boldmath Data analysis}
In the selection of the control sample of $\jpsi \to \pppipi$, only three charged tracks were determined in the event selection, leaving the particle under study as missed.

\begin{center}
\tabcaption{ \label{pppipi_eff_bkg_ana}  Topology analysis when proton is assigned as the missing particle using MC sample of 200M events.}
\renewcommand{\arraystretch}{1.2}
\addtolength{\tabcolsep}{1.5mm}
\begin{tabular*}{80mm}{l@{\extracolsep{\fill}}cc}
\hline \hline
Channel&Events&Ratio\\
\hline

$\jpsi \to \pppipi$&707114&$54.19\%$\\

\multicolumn{3}{c}{Resonances to the same final state}\\

$J/\psi  \rightarrow \bar{p} \pi^{-} \Delta^{++} \to \pppipi$&191565&$14.68\%$\\

$J/\psi  \rightarrow \bar{\Delta}^{--} \pi^{+} p \to \pppipi$&148190&$11.36\%$\\

$J/\psi  \rightarrow \bar{\Delta}^{--} \Delta^{++} \to \pppipi$&129749&$9.94\%$\\

$J/\psi  \rightarrow \bar{p} p f^{'}_{0} \to \pppipi$&33092&$2.54\%$\\

$J/\psi  \rightarrow \bar{p} \pi^{+} \Delta^{0} \to \pppipi$&29619&$2.27\%$\\

$J/\psi  \rightarrow \bar{\Delta}^{0} \pi^{-} p \to \pppipi$&27464&$2.10\%$\\

$J/\psi  \rightarrow \bar{\Lambda} \Lambda \to \pppipi$&13237&$1.01\%$\\

$J/\psi  \rightarrow \bar{p} \rho^{0} p \to \pppipi$&10524&$0.81\%$\\

$J/\psi  \rightarrow \bar{\Delta}^{0} \Delta^{0} \to \pppipi$&5658&$0.43\%$\\

$J/\psi  \rightarrow \bar{p} \omega p \to \pppipi$&2090&$0.16\%$\\

$J/\psi  \rightarrow \bar{p} f_{2}(1270) p \to \pppipi$&898&$0.07\%$\\

\multicolumn{3}{c}{... ...~~~~~~... ...}\\

Sum&$1299200$&$99.6\%$\\
\hline

\multicolumn{3}{c}{Other backgrounds}\\

Sum&$5706$&$0.4\%$\\

\hline \hline
\end{tabular*}
\end{center}

Charged tracks in BESIII are reconstructed from MDC hits. For each charged track, the polar angle must satisfy $|\cos\theta |< 0.93$, and it must pass within $\pm 10$\,cm from the interaction point in the beam direction ($|Z| < 10.0$\,cm), and within $\pm 1$\,cm of the beam line in the plane perpendicular to the beam ($|R_{xy}| < 1.0$\,cm), which is a vertex requirement. The number of charged tracks is required to be at least three.  TOF and specific energy loss $dE/dx$ of a particle measured in the MDC are combined to calculate particle identification (PID) probabilities for pion, kaon, and proton hypotheses. For the three charged tracks, the particle type yielding the largest probability is assigned one by one to each electric charge. The invariant mass of one $p\pi$-pair containing no missing particle is required to satisfy $M_{p\pi}>1.15$ GeV/$c^2$, in order to remove $\Lambda$ particles from decay chains such as $\jpsi \to \Lambda \overline{\Lambda} \to \pppipi$. It is possible some extra tracks remain after the selection, and the influence of extra tracks is discussed in detail in Sec.~{\ref{valid}}(v). Figure~\ref{pt_dis} shows the transverse momentum ($P_T$) distribution of pions and protons from the exclusive MC sample after the above selection criteria for $\jpsi \to \pppipi$.

\section{Tracking efficiency and its uncertainty}

\begin{center}
\begin{figure}
  \centering
  \subfigure{
    \includegraphics[scale=0.12]{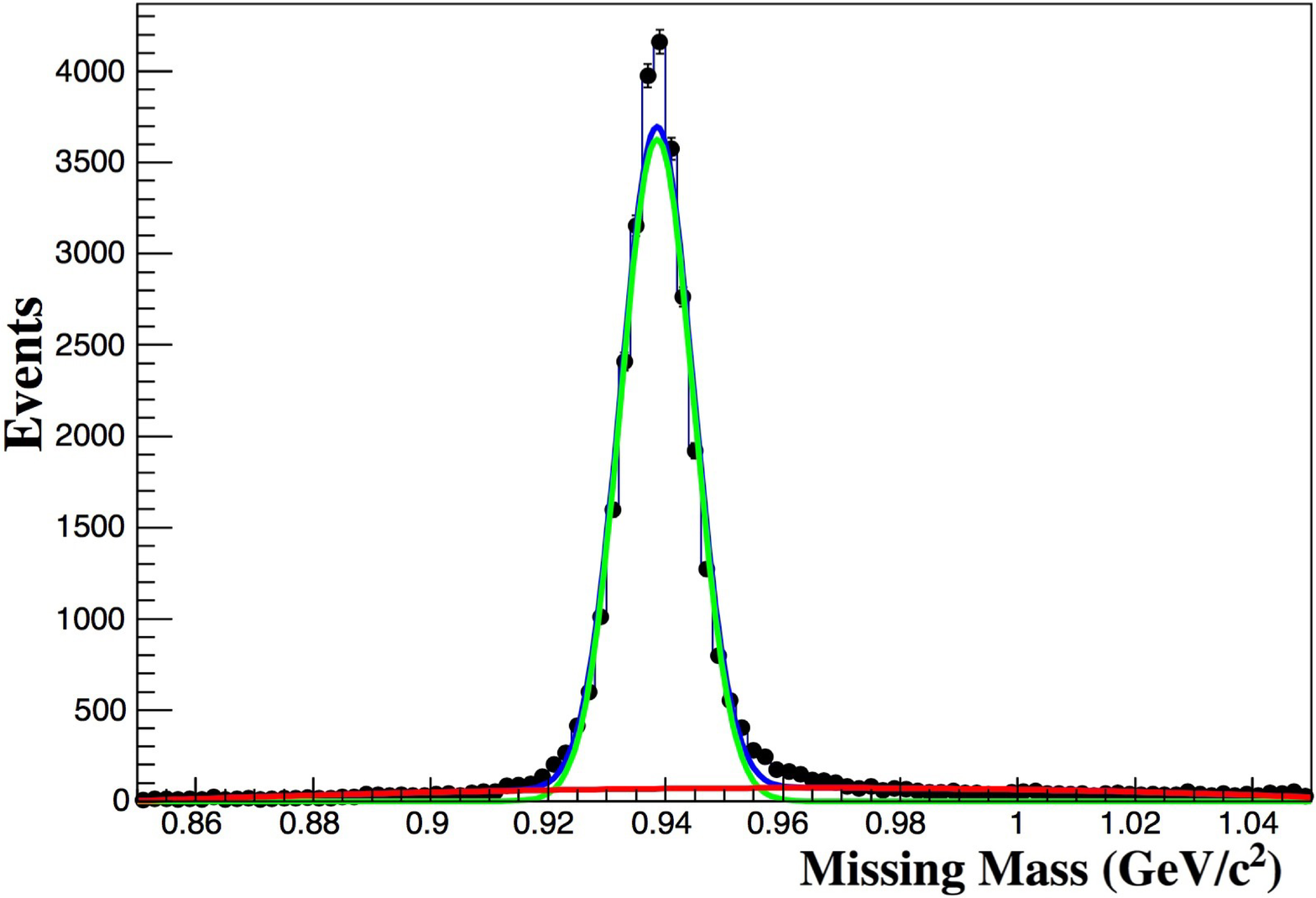}
    \put(-42,115){\bf (a)}}
  \subfigure{
    \includegraphics[scale=0.12]{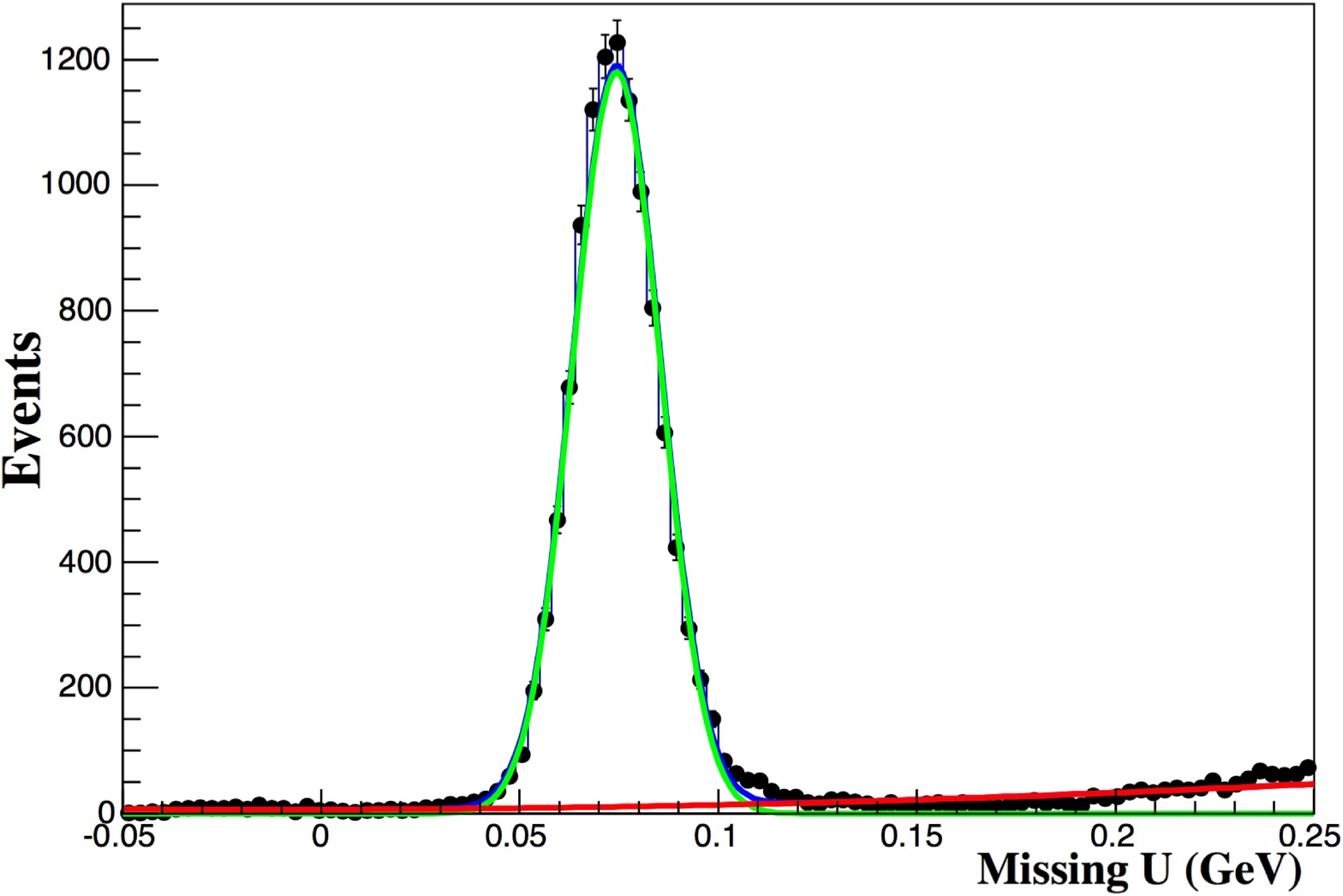}
    \put(-42,115){\bf (b)}}
    \figcaption{\label{trk_eff_fit} (color online) (a) $M_{miss}$ fitting status for protons when $P_T \in (0.3, 0.35)$ GeV/$c$ and $|cos\theta| < 0.7$, and (b) $U_{miss}$ fitting status for pions, when $P_T \in (0.05, 0.1)$ GeV/$c$ and $|cos\theta| < 0.7$, selected from data samples. The green line represents the signal shape described by a Gaussian function, the red line represents the background shape described by a low-degree polynomial, and the blue line is their superposition.}
\end{figure}
\end{center}

 The number of signal events $n$ or $N$ in Eq.~\ref{def_tracking} can be applied by fitting the missing mass spectrum. The $M_{miss}$ spectrum for proton candidates can be described by a Gaussian function and low-degree polynomials, corresponding to signal events and background events, separately. One of the $M_{miss}$ fitting plots from the data sample for protons is shown in Figure~\ref{trk_eff_fit}(a), with the requirements $P_T \in (0.3, 0.35)$ GeV/$c$ and $|cos\theta| < 0.7$ (where $P_T$ and $cos\theta$ are obtained from missing momentum). The distribution of $U_{miss} = E_{miss} - P_{miss}$ (where $E_{miss}$ and $P_{miss}$ represent the energy and momentum of the missing particle, respectively) is applied for pion candidates to achieve the number of signal events, since the momentum resolution will cause a discontinuity at the zero point on the missing mass distribution of pions. The signal and background shape for the $U_{miss}$ distributions of pions are also described by a Gaussian function and  low-degree polynomials. One of the $U_{miss}$ fitting plots for pions from the data sample is shown in Figure~\ref{trk_eff_fit}(b), which requires $P_T \in (0.05, 0.1)$ GeV/$c$ and $|cos\theta| < 0.7$.

The fitting shape used here is inelegant but it is reliable, e.g. the signal peak in Figure~\ref{trk_eff_fit}(a) is not described very well by a Gaussian function, but this has little effect on the efficiency. The fitting status is similar for both $n$ and $N$, thus their ratio, i.e. the efficiency, is insensitive to this minor fitting defect. The number of signal events from the exclusive MC sample can be obtained directly by counting instead of fitting, and the calculated efficiency is consistent with the fitted efficiency (the differences are less than $0.1\%$), which also validates the reliability of the fitting method. Alternative fitting shapes have also been tried, e.g. the signal events described by a double Gaussian function and the background events described by high-degree polynomials, and the differences in efficiencies are less than $0.1\%$. Therefore, a simpler shape, i.e. a Gaussian function with low-degree polynomials, has been chosen for the events description to avoid more parameters in the fitting.

\begin{center}
\begin{figure}
  \centering
  \subfigure{
    \includegraphics[scale=0.35]{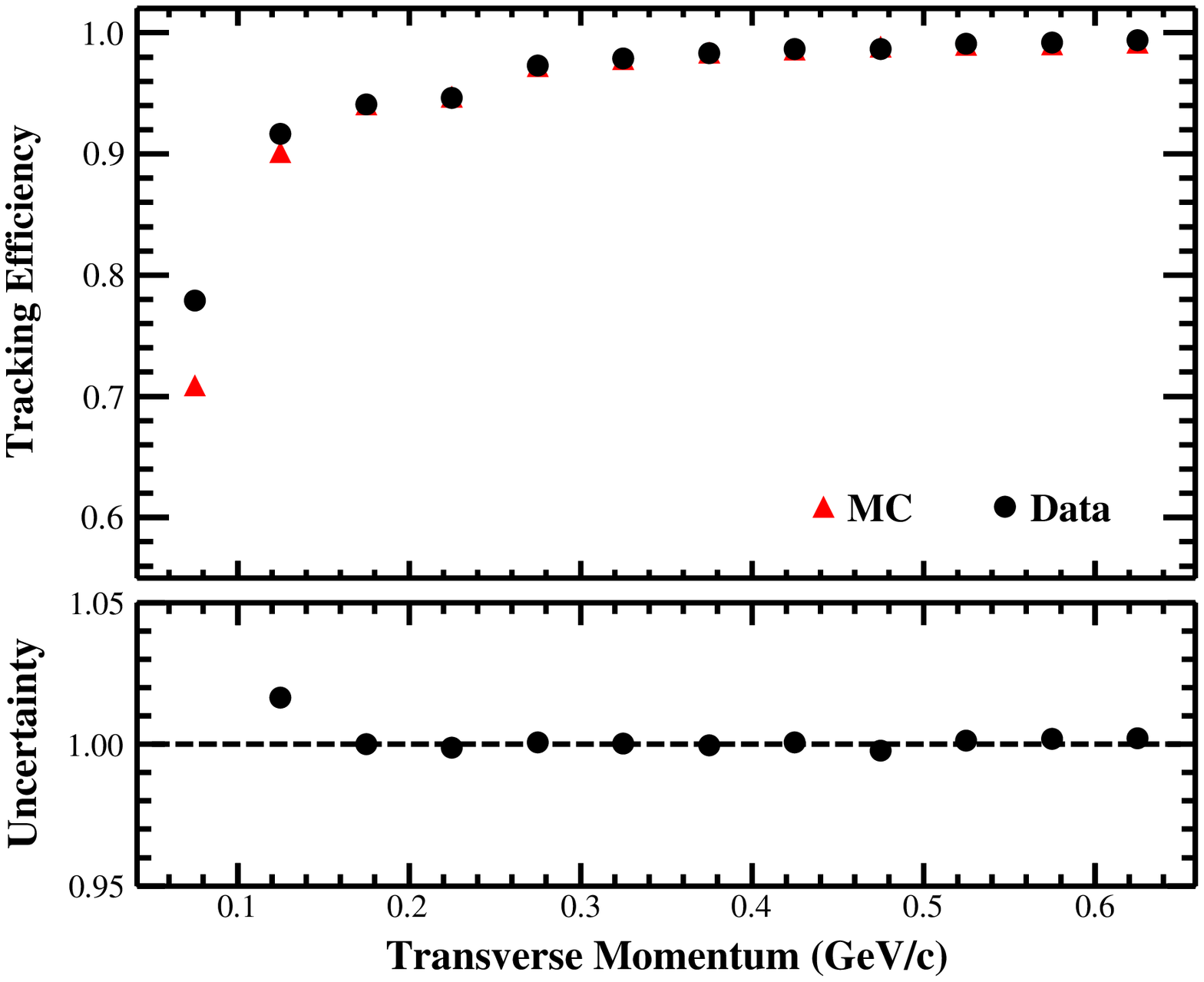}
    \put(-160,75){\bf (a)}}
  \subfigure{
    \includegraphics[scale=0.35]{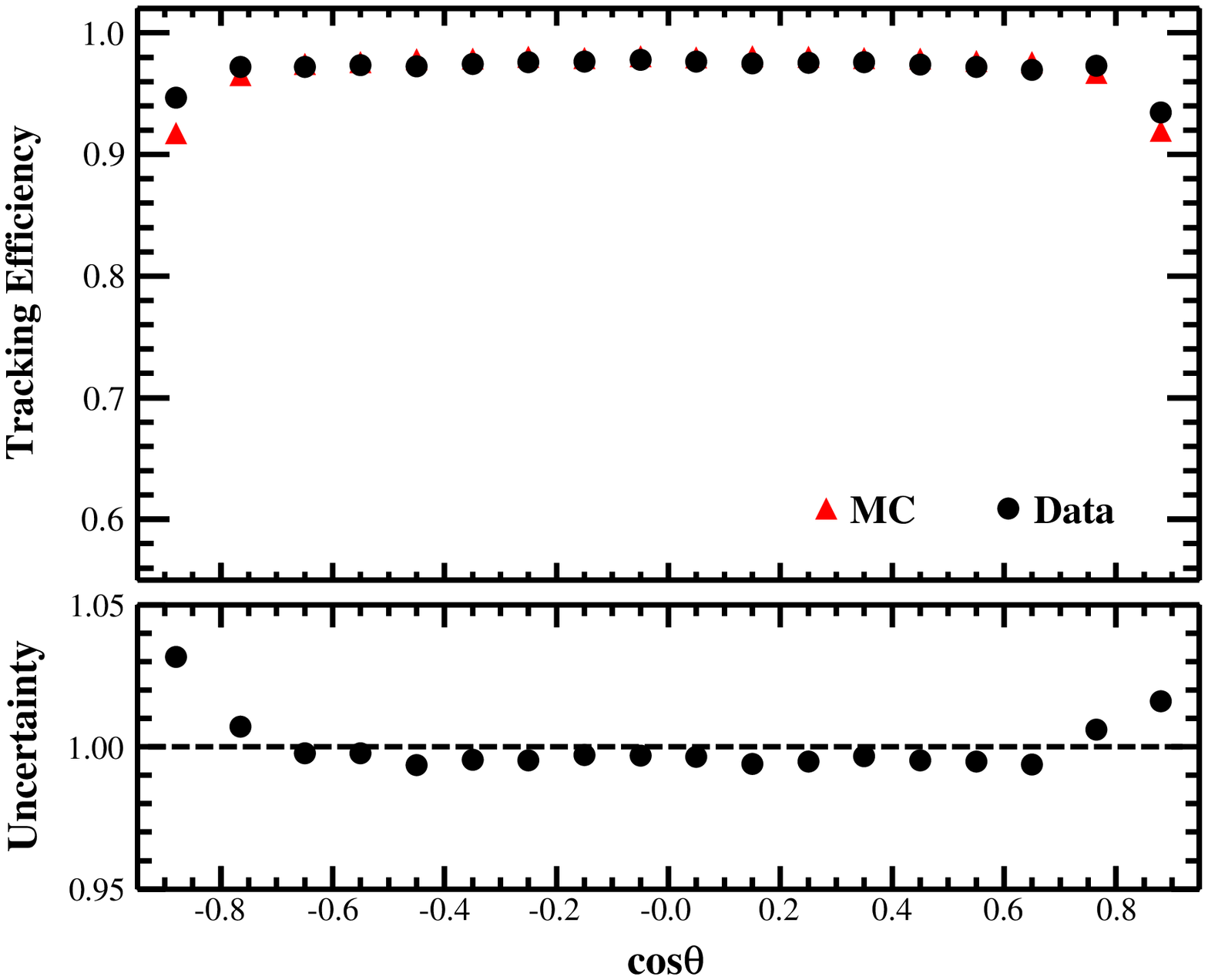}
    \put(-160,75){\bf (b)}}
    \figcaption{\label{trk_eff_09} (color online) One-dimensional tracking efficiencies and their systematic uncertainties for pions (a) with $P_T$, (b) with $cos\theta$ as an example. The red triangles in the upper plots represent the tracking efficiency from the MC sample, the black circles in the upper plots represent the tracking efficiency from the data sample, and the circles in the lower plot represent the corresponding systematic uncertainty.}
\end{figure}
\end{center}

The tracking efficiency and its systematic uncertainty are characterized by $P_T$ and $cos\theta$, since the efficiency is more sensitive to these two variables, which are correlated with the level of track bending and the hit positions of tracks in the MDC respectively. As an example, Figure~\ref{trk_eff_09}(a) and Figure~\ref{trk_eff_09}(b) show the one-dimensional tracking efficiencies and their systematic uncertainties for pions using $P_T$ and $cos\theta$, respectively.

\begin{center}
\begin{figure}
  \centering
  \subfigure[MC]{
    \includegraphics[scale=0.25]{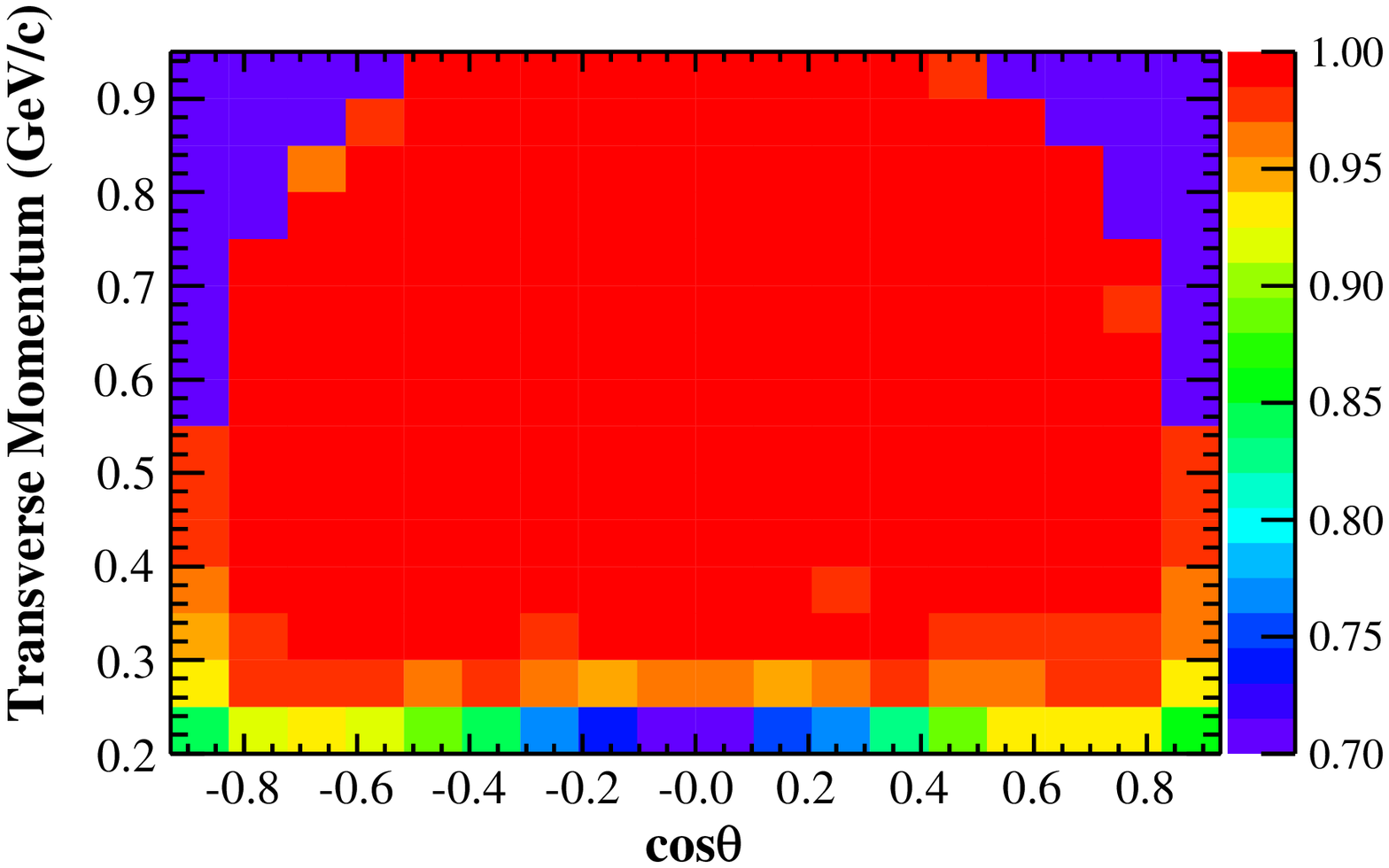}}
  \subfigure[Data]{
    \includegraphics[scale=0.25]{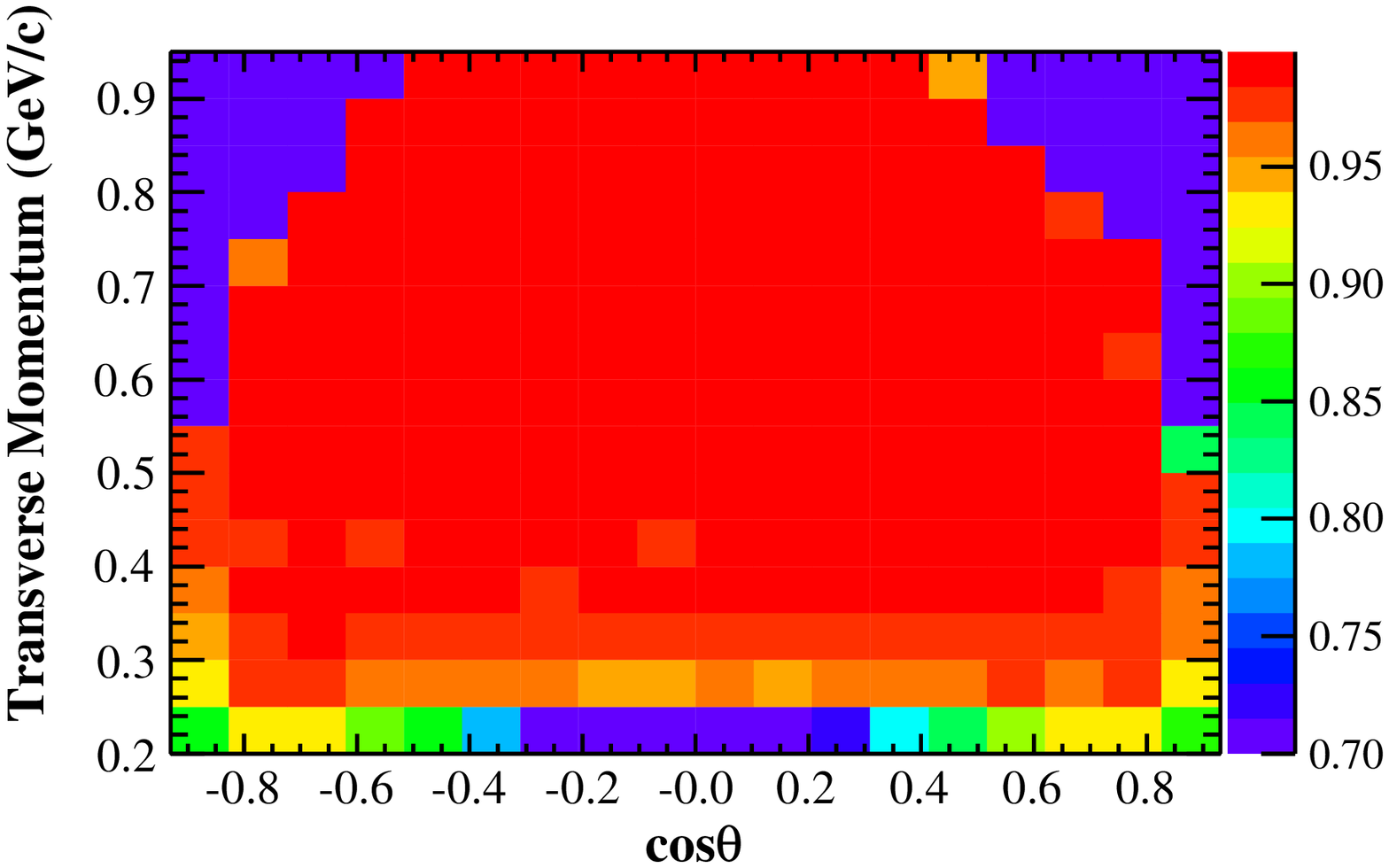}}
  \subfigure[Uncertainty]{
    \includegraphics[scale=0.25]{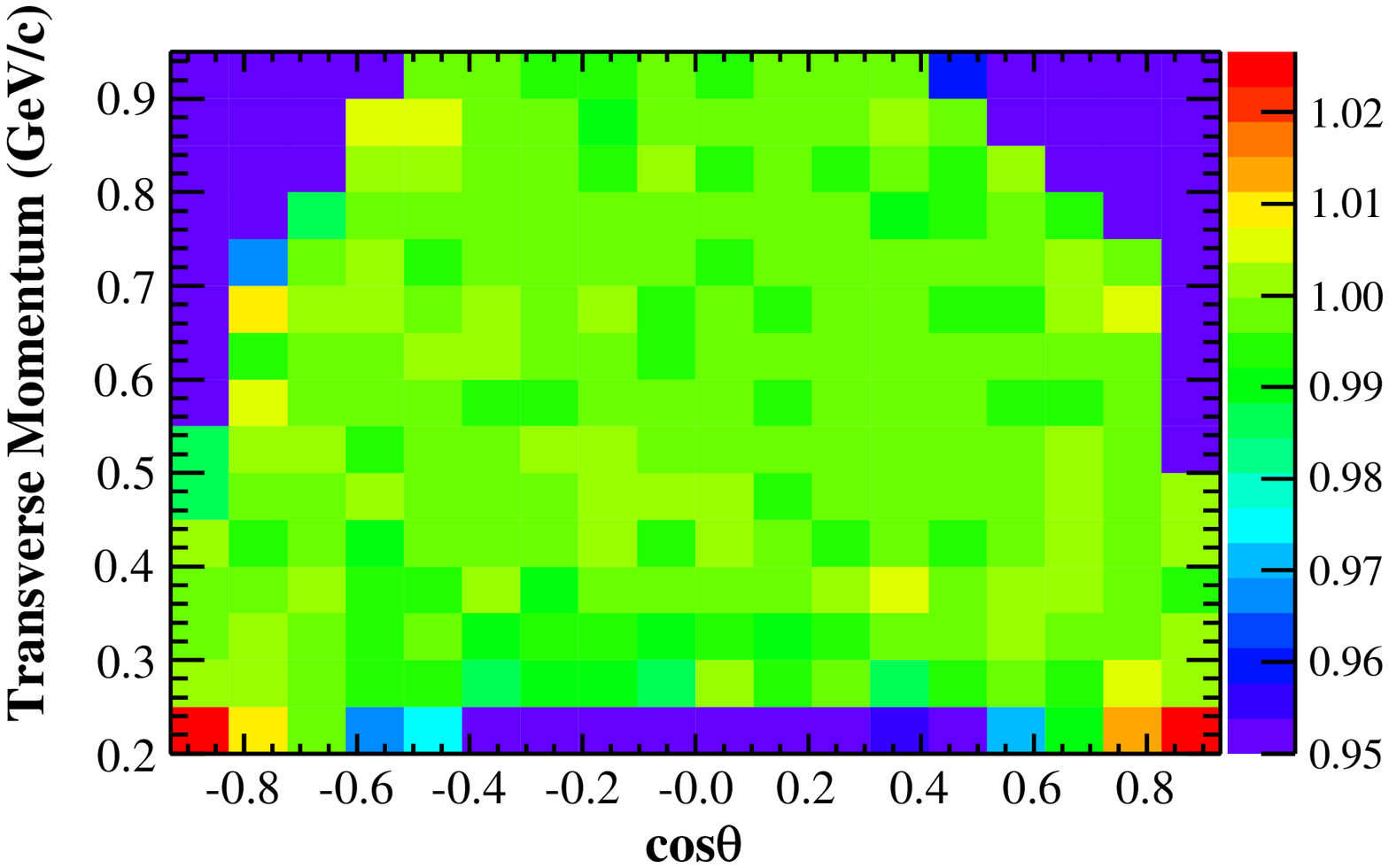}}
    \figcaption{\label{trk_eff_2d_p_09} (color online) Two-dimensional (a) tracking efficiencies from MC sample, (b) tracking efficiencies from data sample, and (c) systematic uncertainties for protons as an example. The x-axis is binned in $cos\theta$  and y-axis binned in $P_T$, while the color of each box representing a corresponding region efficiency.}
\end{figure}
\end{center}

Although different decay processes in physics analyses have different systematic uncertainties for tracking, due to the different distributions of $P_T$ and $cos\theta$, it is possible to provide a general systematic uncertainty for various decay processes. If it is represented in two dimensions, $P_T$ versus $cos\theta$, then the systematic uncertainty for other processes can be obtained by sampling this two-dimensional systematic  uncertainty. An example of such a two-dimensional plot is shown in Figure~\ref{trk_eff_2d_p_09} for protons, as a function of $\cos\theta$  and $P_T$, in which the colors in each box represents the corresponding tracking efficiency or systematic uncertainty.

The results of tracking efficiencies and its systematic uncertainties will be helpful for software development, especially for MDC track reconstruction. For example, one can find the tracking efficiency is less than 0.9 when $P_T < 0.15$ GeV/$c$ in Figure~\ref{trk_eff_09}(a), and the systematic uncertainty is also relatively larger. Low momentum track reconstruction is difficult, since electromagnetic multiple scattering, electric field leakage, energy loss etc. become stronger, and charged tracks with transverse momenta less than 0.12 GeV/$c$ will circle within the MDC. The BESIII software group have used a TCurlFinder method~\cite{curl} to increase tracking efficiency for low-momentum tracks, and improved the MDC MC tuning model to decrease its systematic uncertainty~\cite{tuning}. There is still room for improvement of the low momentum track reconstruction and MC tuning, with the assistance of this study.

\end{multicols}
\begin{multicols}{2}

\end{multicols}
\begin{multicols}{2}

\end{multicols}
\begin{multicols}{2}

\end{multicols}
\begin{center}
\begin{figure}
  \centering
  \subfigure{
    \includegraphics[scale=0.35]{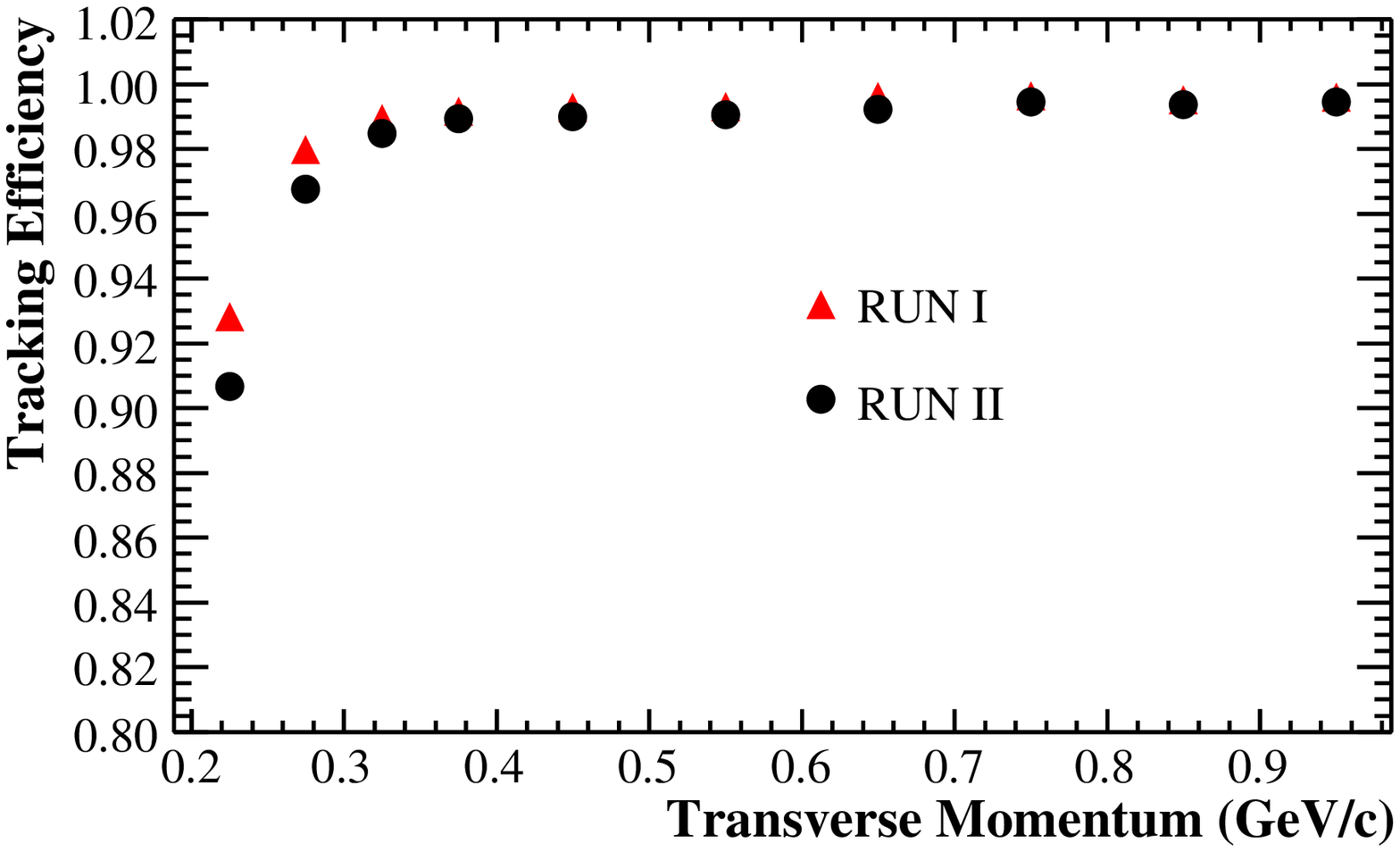}
    \put(-40,30){\bf (a)}}
  \subfigure{
    \includegraphics[scale=0.35]{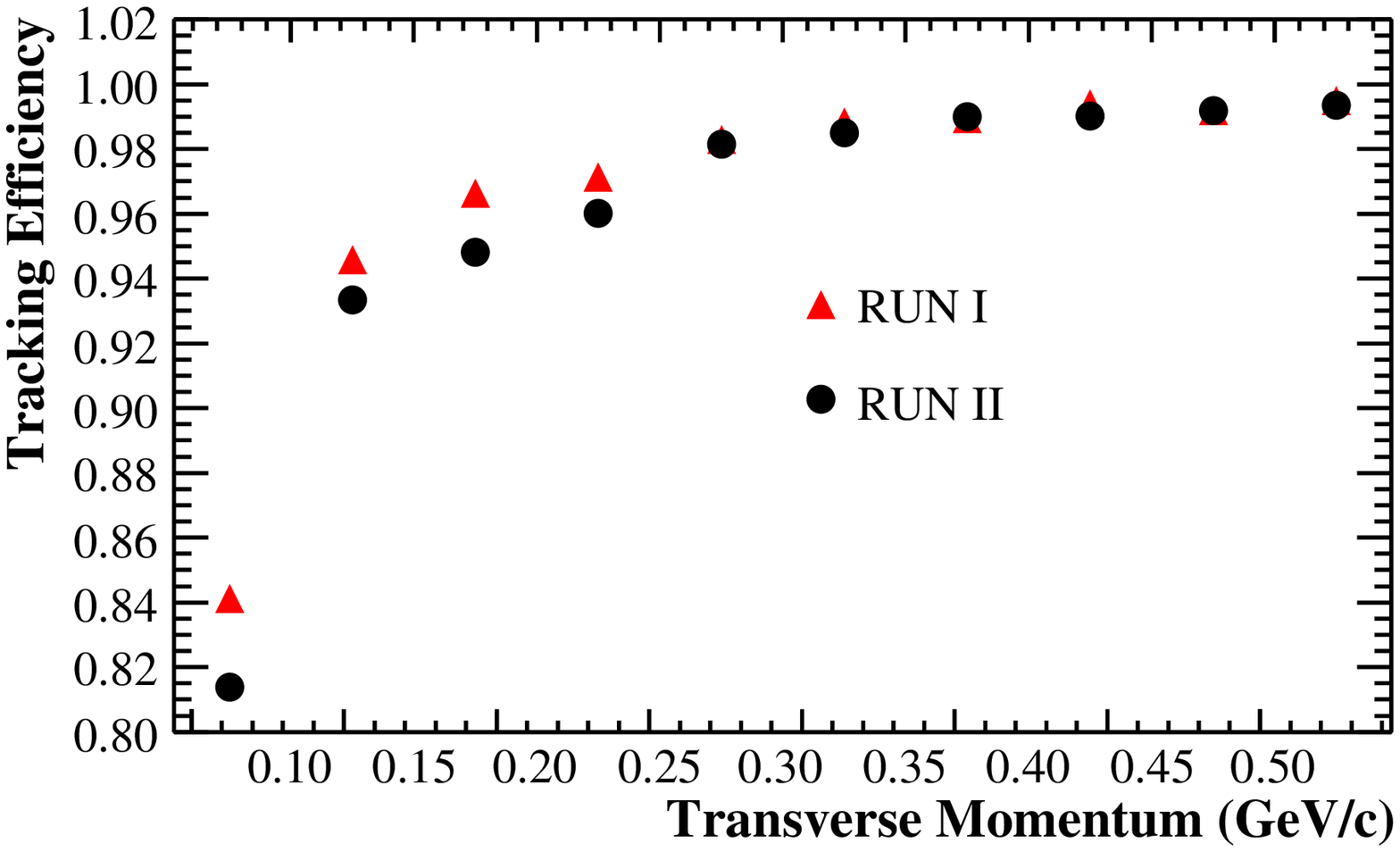}
    \put(-40,30){\bf (b)}}
    \figcaption{\label{noise} (color online) Tracking efficiencies from different noise levels from exclusive MC for (a) proton and (b) pion. The red triangles represent the efficiencies from a typical low noise level run period (RUN I), and the black circles represent the efficiencies from a typical high noise level run period (RUN II).}
\end{figure}
\end{center}
\begin{multicols}{2}

\section{Validation}
\label{valid}
In order to develop a robust methodology for determining the tracking efficiency, a number of essential validations are performed as described below, mainly checking the definition, applicability of results, etc.

(i) Validation of process independent: The tracking efficiency should be process independent, which can be validated by comparing the tracking efficiencies between the single track MC and exclusive MC samples. The single track MC is generated by a virtual process for one pion or proton track with a uniform phase space, and the exclusive MC of the process $\jpsi \to \pppipi$ is produced with a uniform phase space. The momentum and polar angle ranges of the single track MC is required to be similar to the exclusive MC, then a comparison of the tracking efficiency between these two MC samples are supplied. The result shows the tracking efficiencies are consistent between single track MC and exclusive MC samples (the differences are less than $0.1\%$), which is a validation of the process independence.

(ii) Validation of signal extraction: The inclusive MC sample is similar to the data, including both signal and background, and the signal is the same as the exclusive MC. The result shows both the tracking efficiencies and their uncertainties obtained from the exclusive and inclusive MC samples are consistent with each other (the differences are less than $0.1\%$), which proves the signal extraction is reliable.

(iii) Results due to different noise levels: The electronics noise and beam related background at BESIII varies in different run periods during data acquisition, so the effects of different noise levels are considered in this study. Two run periods were chosen to compare the differences in tracking efficiency and uncertainty, one being a typical low noise level run period (RUN I), and the other a typical high noise level run period (RUN II). Figure~\ref{noise} shows the tracking efficiencies with different noise levels from exclusive MC for protons and pions, respectively. The results show that high noise level run period has about $2\%$ lower efficiency compared with the low noise level run period when $P_T < 0.3$\,GeV/$c$.  Since the different noise levels for different periods of data have been introduced into the MC simulation,  the systematic uncertainties of tracking are not subject to the noise level.

\begin{center}
\begin{figure}
  \centering
  \subfigure{
    \includegraphics[scale=0.35]{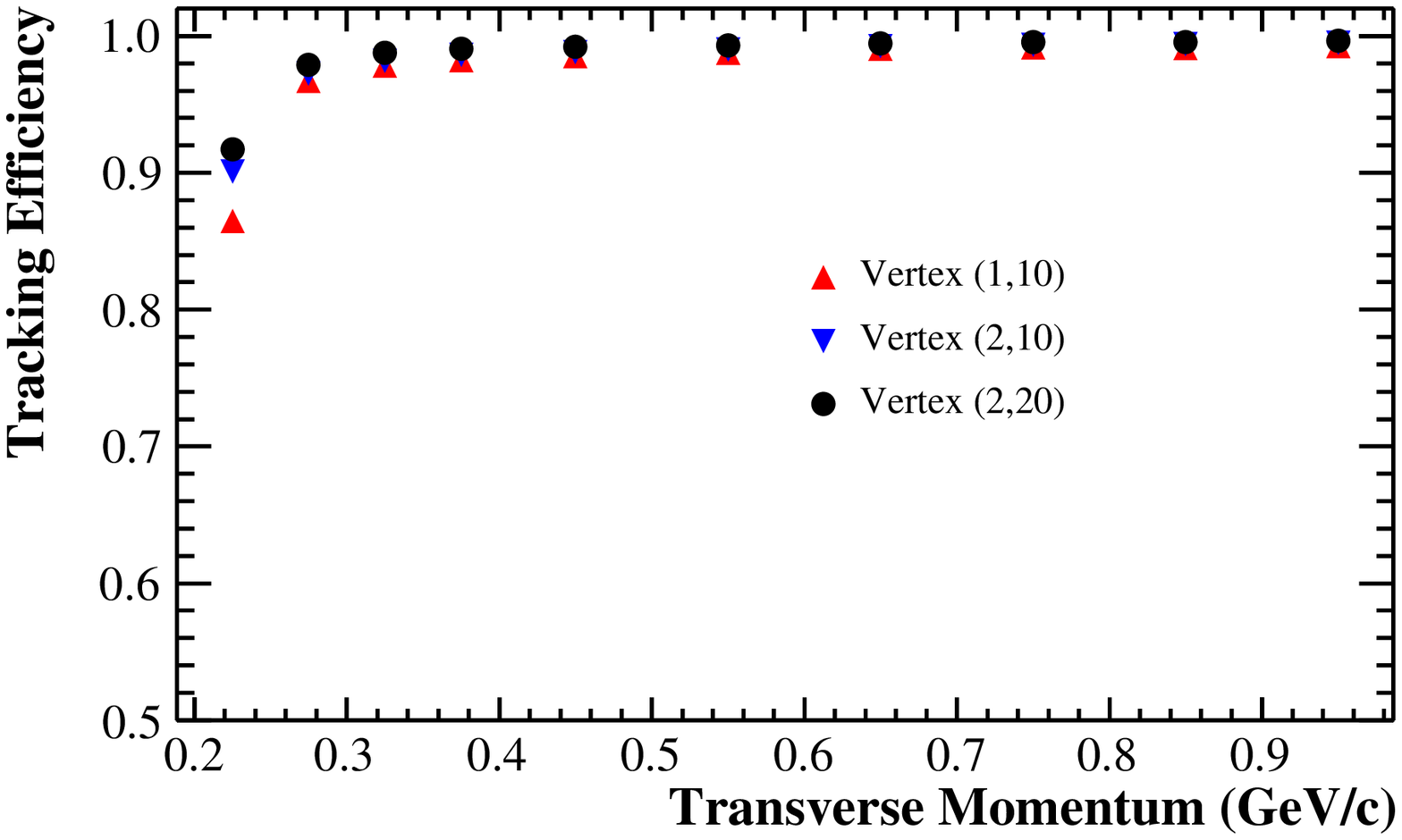}
    \put(-40,30){\bf (a)}}
  \subfigure{
    \includegraphics[scale=0.35]{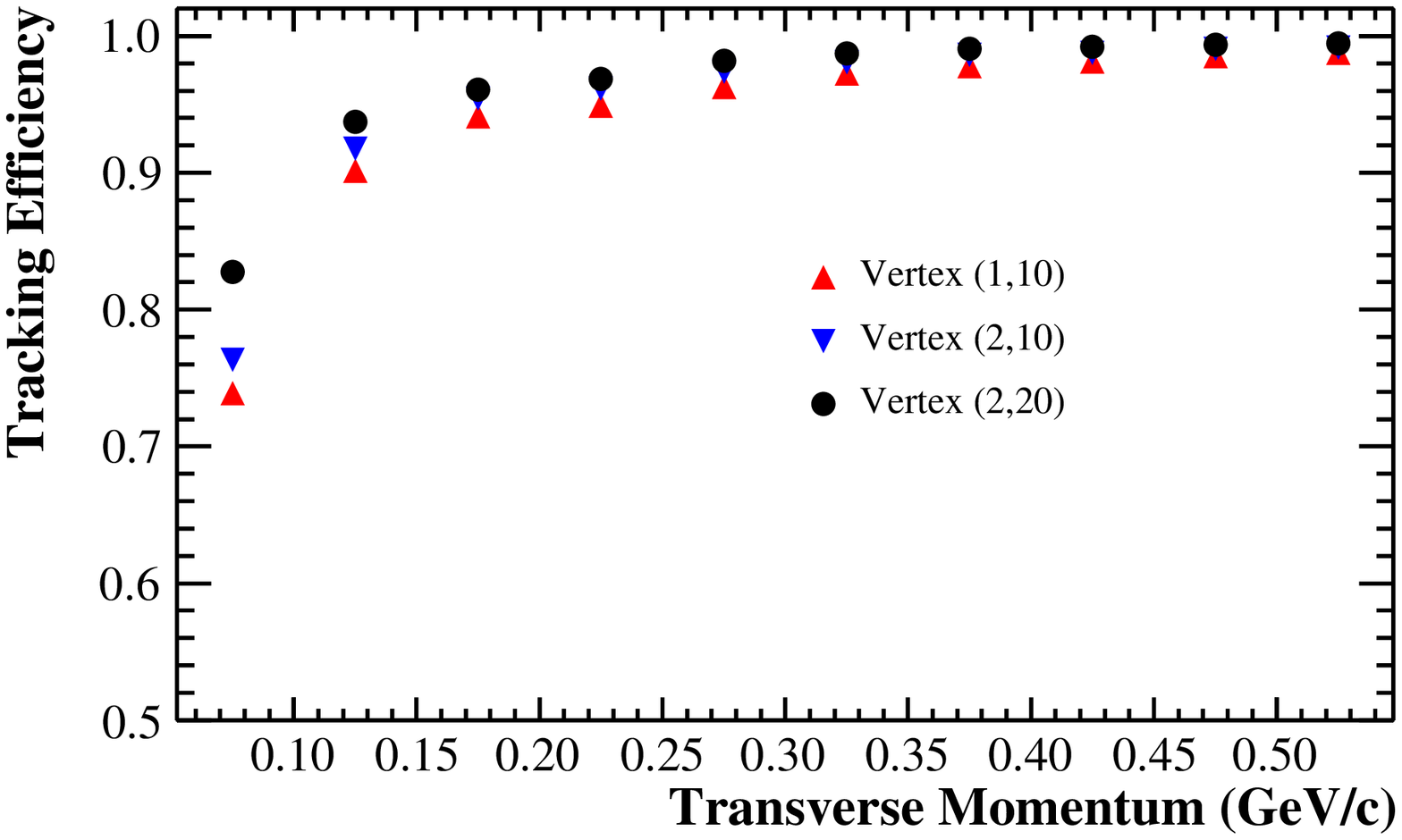}
    \put(-40,30){\bf (b)}}
    \figcaption{\label{vertex} (color online) Tracking efficiencies from different vertex cuts from exclusive MC for (a) protons and (b) pions. The red upward triangles represent the vertex cut $|R_{xy}| < 1.0$\,cm, $|Z| < 10.0$\,cm; the blue downward triangles represent the vertex cut $|R_{xy}| < 2.0$\,cm, $|Z| < 10.0$\,cm; and the black circles represents the vertex cut $|R_{xy}| < 2.0$\,cm, $|Z| < 20.0$\,cm.}
\end{figure}
\end{center}

(iv) Comparing results with different vertex cuts: Although the qualification for  vertex cuts ought to be standard, as an elementary requirement which is  used in almost every analysis, $|R_{xy}| < 1.0$\,cm, $|Z| < 10.0$\,cm; $|R_{xy}| < 2.0$\,cm, $|Z| < 10.0\,$cm; and $|R_{xy}| < 2.0$\,cm, $|Z| < 20.0$\,cm are three common cuts which can be found in different BESIII data analyses. A comparison of these three vertex cuts shows that the efficiencies at low $P_T$ are obviously influenced by the different requirements. Figure~\ref{vertex} shows the tracking efficiencies with different vertex cuts from exclusive MC for protons and pions, respectively. The results show the efficiency from the tightest vertex cut is lower by about $5\%$ compared with the loosest cut at low $P_T$. The data sample behaves in the same way as the MC sample, so the systematic uncertainties of tracking efficiency are insensitive to the vertex cuts.

(v) The influence of extra tracks: The $n$ and $N$ in Eq.~\ref{def_tracking} were explained to be the number of signal events in which the number of reconstructed charged tracks are greater than three and four, respectively, which allows some extra tracks in the definition. A reconstructed track with some extra fake reconstructed tracks was introduced to the tracking efficiency in this study, but the extra tracks may be avoided in most physics analyses, on account of the requirement of zero net charge. The influence of the extra tracks should be evaluated. When count the number of charged tracks in each event, only a few events have the number of charged tracks greater than four, which means the extra fake tracks are seldom reconstructed. The tracking efficiencies with or without the extra tracks were compared for both MC and data samples, and found to have consistent tracking efficiencies (the differences are less than $0.1\%$). Therefore, the influence of extra tracks can be neglected.

\section{\boldmath Summary}
A robust methodology for determining the tracking efficiency and its systematic uncertainty is presented in this paper.  A clear definition of tracking efficiency and its systematic uncertainty was given, a clean sample of $\jpsi \to \pppipi$ was selected and a number of essential validations performed to guarantee the reliability of the method. The tracking efficiency and its systematic uncertainty was determined in both one and two dimensions as a function of both transverse moment and polar angle, which were shown as examples. It is possible to use the general results of tracking efficiency and its systematic uncertainty for the analysis of other processes. \\

\end{multicols}
\begin{multicols}{2}

\end{multicols}

\clearpage

\end{document}